\documentclass[conference]{IEEEtran}
\IEEEoverridecommandlockouts
\usepackage{lmodern}
\usepackage[T1]{fontenc}
\usepackage{cite}
\usepackage{amsmath,amssymb,amsfonts}
\usepackage{algorithmic}
\usepackage{graphicx}
\usepackage{textcomp}
\usepackage{xcolor}
\def\BibTeX{{\rm B\kern-.05em{\sc i\kern-.025em b}\kern-.08em
    T\kern-.1667em\lower.7ex\hbox{E}\kern-.125emX}}

\usepackage{lmodern}
\usepackage[utf8]{inputenx}
\usepackage{indentfirst}
\usepackage{microtype}
\usepackage{tabularx}
\usepackage[nottoc, notlof, notlot]{tocbibind}
\usepackage{hyperref}

\usepackage{booktabs}
\usepackage{multirow}

\newcommand{\NAME}{\textsc{CubeTesterAI}}

\begin{document}

\title{\NAME: Automated JUnit Test Generation using the LLaMA Model}

\author{Daniele Gorla$^1$ \quad
Shivam Kumar$^1$$^,$$^2$ \quad
Pietro Nicolaus Roselli Lorenzini$^2$ \quad
Alireza Alipourfaz$^2$
\\
$^1$University of Rome La Sapienza
\qquad
$^2$PCCube
}

\maketitle

\begin{abstract}
This paper presents an approach to automating JUnit test generation for Java applications using the Spring Boot framework, leveraging the LLaMA (Large Language Model Architecture) model to enhance the efficiency and accuracy of the testing process. 
The resulting tool, called \NAME, includes a user-friendly web interface and the integration of a CI/CD pipeline using GitLab and Docker. These components streamline the automated test generation process, allowing developers to generate JUnit tests directly from their code snippets with minimal manual intervention. 
The final implementation executes the LLaMA models through RunPod, an online GPU service, which also enhances the privacy of our tool.
Using the advanced natural language processing capabilities of the LLaMA model, \NAME\ is able to generate test cases that provide high code coverage and accurate validation of software functionalities in Java-based Spring Boot applications. Furthermore, it efficiently manages resource-intensive operations and refines the generated tests to address common issues like missing imports and handling of private methods. 
By comparing \NAME\ with some state-of-the-art tools, we show that our proposal consistently demonstrates competitive and, in many cases, better performance in terms of code coverage in different real-life Java programs.
\end{abstract}


\section{Introduction}

Software testing is a fundamental component of the software development lifecycle, ensuring that code behaves as intended and meets the desired quality standards. As software systems become increasingly complex, the challenge of maintaining robust and comprehensive test coverage grows exponentially. 
To simplify the task, a common approach is {\em unit testing} \cite{ZH97,R06}, a software engineering activity in which individual units of code are tested in isolation. 

JUnit,\footnote{\url{https://junit.org/}} a widely-used testing framework for Java, plays a crucial role in unit testing by enabling developers to write repeatable and automated tests. However, manual creation of these tests is often labor-intensive, error-prone, and difficult to scale, especially for large and intricate codebases \cite{DF14,GS17}.
However, in the fast-paced world of software development, where rapid iteration and continuous delivery are the norms, there is a pressing need for more efficient, and thus automated, testing methodologies (\cite{LMS99,APX06,NFL06,MBGC22,LKF23,FGG22}, just to cite a few). Traditional automated testing tools, while helpful, often fall short when dealing with complex scenarios, requiring significant manual intervention to ensure adequate test coverage \cite{FSM15}. This gap in existing methodologies highlights the need for innovative solutions that can automate test generation more effectively.

Recent advances in machine learning, particularly in natural language processing (NLP), have opened new avenues for automating various aspects of software development. Large Language Models (LLMs) have shown remarkable capabilities in generating human-like text based on given inputs \cite{TDS22,BG23,GV23,SNET24,BGKJ24,YYG24,TLZL24,WLLJ24,SST24,YLD24,LNR24,CHZ24}. Leveraging these models to automate the generation of JUnit tests presents a promising opportunity to address the challenges of manual test creation, enhancing both efficiency and accuracy in the testing process.

\subsection{Research Problem Statement}

The creation and maintenance of comprehensive JUnit test suites remain a critical bottleneck in modern software development. Manual test generation is usually not satisfactory in situations where individual components or modules depend on each other in intricate ways, including tightly coupled classes, shared resources, and numerous cross-method and cross-class interactions. These interdependencies make it challenging to isolate units for testing, resulting in an increased effort to create accurate and comprehensive tests.

Existing automated tools, like \textit{EvoSuite} \cite{Fraser2011} and \textit{Randoop} \cite{Pacheco2007,PE07}, provide some relief, achieving consistent code coverage and reducing developer intervention, but do not handle intricate code structures \cite{FSM15,FA16}. Moreover, these tools often struggle with scenarios such as private methods or cross-class dependencies, requiring developers to manually address deficiencies that limit the effectiveness of automated testing.
Recent advances in using LLMs (such as ChatGPT, Codex, and StarCoder) for test generation have shown promising  \cite{GV23,SST24,CHZ24}, but continue to face challenges in generating reliable and contextually complete test cases, especially for more sophisticated Java applications. These models often require iterative refinements and may still not meet the coverage standards necessary for production-level code. 


\subsection{Research Objectives}

This paper aims to address these challenges by proposing a solution for the automated generation of JUnit tests using the LLaMA model \cite{LLAMA}, a family of autoregressive large language models released by Meta AI starting in February 2023. The latest version is LLaMA 3.2, released in September 2024.

The specific objectives of this research are:

\begin{description}
    \item[RO1] \textbf{Integrate the LLaMA model into a testing framework} to automate the generation of JUnit tests for Java applications, focusing on improving the coverage and accuracy of the tests.
    \item[RO2] \textbf{Optimize and enhance the performance of the LLaMA model} using various techniques that ensure that the generated tests are reliable, executable, and provide meaningful insights into code functionality.
    \item[RO3] \textbf{Conduct a comparative analysis of the generated tests} to evaluate their quality, coverage, and effectiveness, mostly when compared with state-of-the-art tools that face the same research problem as ours.
\end{description}
In carrying on these objectives, we also aimed at developing a user-friendly web interface that allows developers to easily generate and refine JUnit tests based on their code snippets; this reduces the manual efforts required in the testing process. Furthermore, we also aimed at implementing an automated Continuous Integration/Continuous Deployment (CI/CD) pipeline using GitLab and Docker, which triggers code changes, generates JUnit tests, validates them, and integrates them seamlessly into the development workflow.

\subsection{Outcome of the Research}

In this paper, we present \NAME, a tool that fully achieves the RO1 and RO2 mentioned above and is able to automate the generation of the JUnit tests for Java applications using the Spring Boot framework.  \NAME\  leverages the LLaMA model to improve the efficiency and accuracy
of the testing process, includes a user-friendly web interface and the integration of a CI/CD pipeline using GitLab and Docker. These components streamline the automated test generation process, allowing developers to generate JUnit tests directly from their code snippets with minimal manual
intervention. Furthermore, \NAME\ executes  the LLaMA models through RunPod \cite{runpod2024}, an online GPU service, which also enhances the privacy of our tool. Indeed, by hosting the model on a dedicated server, we ensure greater control over data handling processes, thereby mitigating potential risks associated with the use of third-party APIs. 

To improve the accuracy and reliability of generated tests, we also implemented a feedback loop: when a generated test method fails,  \NAME\  provides LLaMA with the specific test method content and the associated error stack trace. LLaMA then refines the test method until it is passed or the maximum number of iterations specified by the user is reached. 

A notable feature of \NAME is the possibility of testing private methods or methods that use private methods. This was a significant challenge, since such methods cannot be directly mocked, leading to difficulties in testing. Our approach consists in testing them indirectly, while testing the public methods that invoke them. This has the advantage of preserving encapsulation or mocking constraints while ensuring high code coverage.

As we expected, automating the generation of JUnit tests using the LLaMA model can significantly reduce the time and efforts required for testing, allowing developers to focus more on innovation and less on the repetitive task of test creation. 
This became evident while working for achieving RO3. To this aim, we performed several tests, first to discover the best LLaMA model to use, then to show that LLaMA performs better than ChatGPT, and finally to see how  \NAME\  performs compared with some state-of-the-art tools.

In detail, we first considered four LLaMA models, namely 
\texttt{meta-llama/Llama-2-7b-hf}, \texttt{meta-llama/Llama-\linebreak 2-13b-hf}, 
\texttt{codellama/ CodeLlama-34b-hf}, and
 \texttt{gradientai/Llama-3-70B-Instruct-Gradient-\linebreak 1048k}. As expected, we discovered that the latter model outperforms the others, in terms of number of generated tests, passed tests (both in the first and in successive iterations), and code coverage. For this reason,  \NAME\  includes this latter model.
Moreover, a detailed cost analysis revealed that processing a Java class of approximately 100 lines with three methods incurs a cost of around 4€; this cost only depends on the GPU usage on RunPod, which is approximately 20 minutes and involves 53 requests to RunPod. This underscores the practicality and affordability of using cloud-based GPU services for automated test generation.

Then, we tried to integrate ChaptGPT in place of LLaMA in our tool and see how the results change. It turns out that moving from LLaMA to ChatGPT reduces the ability of \NAME, in terms of both generated tests (168 vs. 185), test passed (114 vs. 152) and code coverage (76.3\% vs. 88.6\%), when run on three real-life Java programs. This fully justifies the use of LLaMA in \NAME.

To conclude, we compared \NAME\  with some state-of-the-art tools and on existing datasets, to comparatively assess its performance. 
The first comparison was against the tool \texttt{ChatUniTest} \cite{CHZ24}, which focuses on generating unit tests using LLMs for unseen projects. By running our tool on the datasets used by them, we discovered that  \NAME\  produces a much higher average coverage (86\% vs. 59.6\%). 
Next, we compared our results with those reported in \cite{SST24}. This study tested multiple models, including \texttt{GPT-3.5-Turbo} \cite{gptTurbo}, \texttt{StarCoder} \cite{starcoder}, \texttt{Codex-2K} and \texttt{Codex-4K} \cite{codex}, and \texttt{EvoSuite} \cite{FA16}, on the HumanEval dataset \cite{HumanEval}.  \NAME\  achieved the highest code coverage of 98.18\%, outperforming all the models tested in the study.
Finally, we evaluated \NAME\  against the tools reported in \cite{GV23}, a study that investigates ChatGPT's unit test generation capabilities on a dataset of 33 Java programs taken from \cite{AR21}.  \NAME\  achieved a competitive overall coverage of 97\%, which surpasses the ChatGPT tool (90.2\%) and is close to the Baseline Suite (99.5\%), a set of tools that includes \texttt{EvoSuite} \cite{FA16}, \texttt{Randoop} \cite{PE07}, \texttt{Palus} \cite{Z11}, and \texttt{JTExpert} \cite{SPG15}.


\subsection{Related Work}

Automated test generation has been an active research topic within the software engineering community for several years. The progression from traditional test automation tools to leveraging advanced machine learning models (like LLMs) has brought significant improvements in testing efficiency and accuracy. 



\subsubsection{Machine Learning in Software Testing}
The introduction of machine learning models has marked a significant evolution in software testing. Researchers have explored using \textit{Long Short-Term Memory (LSTM)} networks \cite{LSTM} and \textit{Transformer-based} sequence-to-sequence models \cite{vaswani2017attention} to generate unit tests from code snippets or descriptions \cite{Tufano2020}. However, these models faced limitations in maintaining the semantic integrity of code, particularly with long sequences, thus failing to produce consistently high-quality tests.

More recently, \textit{transformers} (like \textit{BERT} \cite{devlin2018bert} and \textit{GPT-3} \cite{Brown2020}) have been applied to test generation due to their ability to understand both code and natural language. \textit{BERT} has shown strengths in extracting features and relationships from code, while \textit{GPT-3} (and other autoregressive models like \textit{Codex} \cite{codex}) have demonstrated success in generating syntactically correct and contextually relevant unit tests \cite{Austin2021}. 

\subsubsection{Large Language Models for Test Generation}
The application of LLMs like \textit{GPT-3}, \textit{Codex}, and \textit{LLaMA} for automated test generation is a developing field. \textit{ChatGPT}, \textit{Codex}, and similar models have shown potential in the generation of automated unit tests but often face challenges related to consistency in output and dealing with complex class dependencies \cite{SST24,CHZ24}. Studies have demonstrated that these models generally provide high coverage for basic cases but require iterative refinement for more intricate scenarios, which can become a bottleneck in the test generation process \cite{YLD24}.

\NAME\ addresses these issues through the integration of pre-processing, post-processing and the use of the LLaMA model, which is specifically designed to handle complex code semantics and generate highly accurate test cases. Compared to other LLMs, LLaMA can handle longer contexts while maintaining coherence across longer sequences; this makes it superior in tasks that require in-depth contextual understanding of code, when compared to other models (like \textit{StarCoder} and \textit{CodeT5}) for automated test generation.

\subsubsection{Iterative Refinement Techniques in Test Generation}
Several tools use iterative refinement to improve robustness of the tests generated. For example, \textit{EvoSuite} \cite{Fraser2011} generates tests that focus on the implemented behavior of the unit under test, rather than the expected behavior. The tests generated by EvoSuite pass by construction, as they are derived to align with the current implementation. Although this approach ensures initial correctness, it does not necessarily validate the correctness of the software against its intended specifications \cite{FSM15}. \textit{ChatUniTest} \cite{CHZ24}, a tool leveraging GPT-based models, generates test cases and refines them based on feedback; however, it struggles to achieve high coverage consistently due to the context limitations of the underlying LLM.

\NAME\ incorporates a sophisticated feedback loop that uses error logs from failed test cases to iteratively refine them until they pass or reach a maximum number of iterations. This method significantly reduces the manual effort required compared to previous refinement strategies \cite{YLD24} and provides a higher reliability in the generated test suites.




\subsubsection{Integration of CI/CD in Test Generation}
Integrating test generation tools into a CI/CD pipeline has become crucial for modern software development practices. Traditional tools like \textit{Jenkins} \cite{Mohan2018} provide basic automation capabilities but lack seamless integration for automated test generation without significant manual scripting. \textit{TestSpark} \cite{Smith2023} and other tools have started to incorporate LLMs into the pipeline, but they often face challenges related to model deployment and scalability.

\NAME\ leverages GitLab CI/CD and Docker for effective pipeline integration. By automating test generation, validation, and deployment, \textit{ \NAME\ } ensures that tests are generated and refined continuously as code evolves, thus reducing the manual burden on developers. 

\subsubsection{User Interface Aspects in Automated Test Tools}
The user interface plays a crucial role in the usability of test generation tools. Traditional automated tools, like \textit{EvoSuite} \cite{Fraser2011} and \textit{Randoop} \cite{Pacheco2007,PE07}, are primarily CLI-based, limiting their accessibility to less technical users. In contrast, \NAME\ features a web-based interface with chatbot capabilities, allowing developers to interact directly with the LLaMA model to generate and refine test cases in a more intuitive way and to enable a real-time iteration.

\section{\NAME}
\label{chap:3} 

\begin{figure*}[t]
    \centering
    \resizebox{.9\textwidth}{!}{%
        \includegraphics[width=0.8\linewidth]{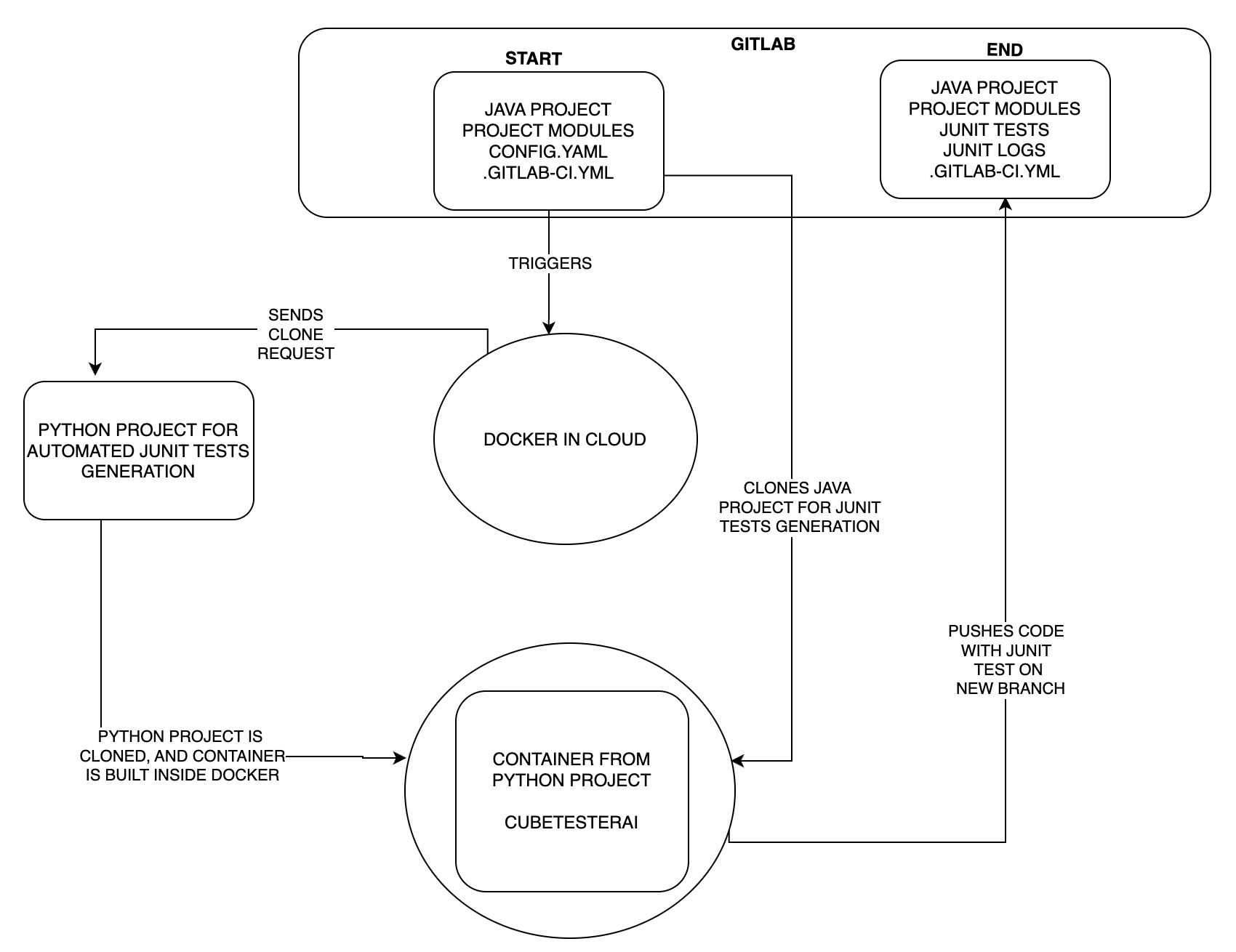}
    }
    
    \caption{Workflow of \NAME\ in the CI/CD pipeline, where:
    \textbf{GitLab} represents the starting and ending point of the Java project with configuration files and generated tests;
    \textbf{Docker in Cloud} creates a container by cloning the Python project; and
    \textbf{Python Project (\NAME)} is responsible for cloning Java projects and JUnit test generation.
    Arrows indicate the flow of actions between different components.}
    \label{fig:architecture}
\end{figure*}

\subsection{Methodology}

%
%
The research design underlying \NAME\ was predominantly qualitative, incorporating the principles of user-centered design to enhance the relevance and usability of the automated test generation process. 
The Test-Optimized Development Model (TODM), derived from User-Centered Design principles, was specifically created to guide the development of \NAME. 
We started by identifying the challenges (private methods, high code coverage, and dependency management). We then moved to AI-centric prototyping, leveraging LLaMA models to validate test generation capabilities and refine outputs iteratively. We introduced a feedback loop to enable continuous improvement, where test failures feed directly into the model for adjustment, reducing manual intervention. After this, we moved to the integration phase, to improve scalability, with a CI/CD pipeline built on GitLab and Docker, supported by RunPod for efficient resource management. Finally, we introduced user-centric enhancements to ensure accessibility through a web-based interface.

%
%

A critical component of the implementation is the iterative testing and refinement of the automated process, following the UCD principle of continuous user feedback and involvement.
Furthermore, our implementation ensures scalability of the model, optimizes performance, and addresses ethical concerns. Specifically, data privacy and security is paramount when making API calls to a dedicated server that hosts the LLaMA model. By hosting the model on a dedicated server, \NAME\ ensures great control over data handling processes, thus mitigating potential risks associated with third-party API usage.
%
The key tools and technologies used in \NAME\ include Docker (for containerization), GitLab CI/CD (for pipeline orchestration), and LLaMA models (for natural language processing). 

The LLaMA model is integrated into our testing framework as shown in Fig.~\ref{fig:architecture}.
When the \texttt{config.yaml} file is modified, the pipeline (\texttt{.gitlab-cl.yml}) triggers a Docker in the cloud. This Docker sends a clone request to the Python project repository, clones the Python project repository and creates a Docker container from it. Then, this container clones the Java project for automated JUnit test generation and, after generating and validating JUnit tests, it pushes the Java code to a new branch on gitlab.

The \texttt{config.yaml} contains crucial information such as the Java version, the class name for which JUnit tests are to be generated, the specific methods (if any) for which tests are needed, and the maximum number of iterations for refining test methods that do not pass.
In the pipeline file, users are instructed to trigger our Docker image (\texttt{docker-junit-test}) with the repository URL and the branch name as parameters. This ensures that our system has access to the necessary data in the repository for test generation.
%
Upon receiving the trigger, \texttt{docker-junit-test} pulls the \NAME\ Python project repository (which contains the entire logic for test generation) and builds a Docker image (using the Python project repository pulled) that runs, using the repository URL and the branch name provided as parameters.

The next step involves cloning the main Java project from the specified branch using the  parameters provided. Once cloned, the system reads the \texttt{config.yaml} file to gather the necessary configuration details.
%
Based on them, \NAME\ prepares data. In particular, it identifies the class name and method names (if specified) from the configuration; it extracts all method names from that class (if only the class name is provided); it extracts, for each method, the method content, the autowired components (Spring \texttt{@Autowired} facilitates automatic dependency injection, promoting loose coupling and easier testing), the import statements and the package name;
it identifies and extracts all object dependencies related to the method, such as entities (that define the data model and are mapped to database tables, managed by ORM frameworks) and DTOs (that are used to transfer data between layers or systems, enhancing encapsulation, security, and flexibility).

The prepared data are then used to create prompts for the LLaMA model. The prompts are designed to instruct the model on generating comprehensive unit tests for Java methods using JUnit 5 conventions. Among other things, the prompt includes the Java version,
instructions to generate comprehensive tests (for 100\% code coverage),
and details on the method content, entity classes, DTOs, and specific naming strategies (see Section \ref{sec:prompts} for more details on the prompts).

Upon receiving the generated tests from the model, \NAME\ saves them in a temporary file; 
each test method is then run individually using Maven, and logs are stored.
If a test fails, \NAME\ refines the test by resending to the model the test method content and the relevant logs, following an iterative process, until the test passes or the maximum number of iterations is reached.
If the maximum number of iterations is reached and the test still does not pass, \NAME\ saves the final version of the test along with the error log, and pushes it to GitLab at the end of the process. This ensures that the user can manually correct the test. Indeed, in most cases where the tests do not pass (even after the maximum number of iterations), the errors are minor and can be easily resolved manually.

After processing all classes and methods specified in the configuration file, \NAME\
pushes the changes to a new branch named \texttt{current-branch-junit-tests- timestamp}, 
cleans the Docker volume and removes the Docker image to free up resources.


\subsection{Prompt Creation and Evaluation}
\label{sec:prompts}

The design of prompts was pivotal in ensuring high-quality, comprehensive JUnit tests with maximum code coverage. The prompts are crafted by adhering to JUnit 5 conventions and Mockito guidelines. The key aspects of the prompt include:
\begin{itemize}
    \item \textbf{Task Definition:} The prompts clearly specify the goal of achieving 100\% code coverage, using JUnit 5 and Java version \texttt{\{java\_version\}}.
    \item \textbf{Mocking and Setup:} Instructions to use \texttt{@ExtendWith(MockitoExtension.class)}, \texttt{@InjectMocks}, and \texttt{@InjectMock} are explicitly provided while avoiding \texttt{@BeforeEach} setup methods.
    \item \textbf{Context Details:} The class name, method body, external dependencies such as DTOs and ORM entities, and edge cases are included to effectively guide the model.
    \item \textbf{Naming Conventions:} Test methods follow the Given-When-Then strategy, and the test class is named as \texttt{\{class\_name\}+Temp}.
\end{itemize}
We report here a sample prompt:

\fbox{\begin{minipage}{23em}
As a seasoned Java developer, I need your expertise in crafting comprehensive unit 
tests for a given Java class encapsulating complex business logic.\\
- Follow the Given-When-Then naming strategy for each test method.\\
- Use JUnit 5 only for writing assertions.\\
- Consider all potential edge cases and boundary conditions.\\
- Name the test class “{class\_name}+Temp."\\
- Use the spring-boot-starter-test dependency.\\
- Here is the code to be tested: {method\_body[0]}.
\end{minipage}}

\medskip

The prompts are iteratively refined based on error logs and coverage reports. This process ensures that the tests generated meet high quality and robustness standards, reducing the need for manual adjustments.

\subsection{Challenges and Solutions}

Throughout the integration and implementation process, several challenges were encountered:
\begin{enumerate}
\item {\em Log Size Management:}
 The logs generated by running Maven tests were often too large, exceeding the context window of the LLaMA model. Hence, we extracted only the error lines from the logs. This significantly reduces the size of the logs while retaining the critical information needed for debugging and refining the tests.

\item {\em Handling Compilation Errors:}
During the refinement process, if a compilation error occurred in a test method different from the one currently being tested, it was difficult for the model to correctly identify and address the error.
To isolate errors, we saved all the tests generated by LLaMA in a temporary text file and then extract one test method into a separate new java file. This approach ensures that each test method is compiled and tested independently, making it easier to pinpoint and correct specific errors through iterative refinement.

\item {\em Ensuring Accurate Refinement:}
To ensure that the model accurately refines the tests based on specific errors and feedback, we check that \NAME\ has the necessary context to improve the test methods effectively, by providing detailed error logs and maintaining an iterative refinement process.

\item {\em Resource Management:}
Running resource-intensive models like LLaMA on local Docker instances could be inefficient and slow. For this reason, we implemented an API call to a LLaMA model hosted on Runpod \cite{runpod2024}, a cloud server, to ensure efficient model execution. This approach allows us to leverage cloud resources for better performance while maintaining the security and privacy of the code being tested.

\item {\em Managing Temporary Files and Reducing Compilation Errors:}
For each unit test generated by the application, a separate temporary file is created to have the unit test inside. This methodology was chosen to avoid compilation errors. If all unit tests are placed in the same file, during the monitoring of running a test, there is a possibility of having a compilation error from another test that could affect the monitored test. In this case, the process of test refinement is complex. In order to reduce the complexity, each test is monitored in a separate file; then, after all the refinements are done, all the successful unit tests are combined inside a single file. The next step is to build the whole project. In this situation, the unit tests files that are not successful even after the iteration refinement is completed will be converted to a text file, to be fixed by the developer.
\end{enumerate}


\subsection{Web Interface Development}
\label{chap:5} 

The web interface was designed to assist users in generating JUnit tests and in making necessary modifications through an interactive chat-bot system. This interface allows users to input their requirements and receive test code suggestions from the LLaMA model, enabling efficient test generation and refinement.
The decision to develop a web-based interface stems from the observation that LLaMA-generated tests often cover most edge cases and have minimal logical errors regarding the main method. However, minor errors, such as missing imports or incorrect data types, are common. By allowing users to directly interact with the model through the web interface, we allow them to request specific changes, saving time and efforts. Indeed, users no longer need to think about edge cases or write extensive code manually; instead, they can focus on making minor adjustments, which the model handles efficiently through the chat-bot interaction.

The backend of the web interface was developed using Flask, a lightweight and efficient web framework for Python. 
Due to privacy concerns associated with the transmission of sensitive data over external APIs, the API calls were implemented using Runpod \cite{runpod2024}, where LLaMA runs locally. This ensures that all data remain secure and private.

To handle multiple users that interact simultaneously, performance optimization strategies were employed. These include load balancing, efficient database queries, and caching mechanisms to ensure that the web interface runs smoothly and efficiently.





\subsection{Automated Pipeline with GitLab and Docker}
\label{chap:6}

\NAME\ aims at integrating the test generation system directly into the developer's workflow. Since GitLab is a common repository where developers maintain their updated code, it was an ideal choice for implementing our CI/CD pipeline.
%
%
The pipeline was configured to trigger only when changes are detected in the \texttt{config.yaml} file; this ensures that the pipeline runs only when necessary.
%
The GitLab pipeline triggers a custom runner, which is set up on the cloud, to ensure a consistent and always-available environment, and by using Docker to manage the environment.
%
%

Some key benefits of using Docker in this context include:
\textbf{Consistency:} Docker ensures that the environment is the same across different runs, eliminating issues caused by discrepancies in dependencies or configurations;
\textbf{Isolation:} Each pipeline run is isolated within its own Docker container, preventing conflicts between concurrent runs and ensuring a clean state for each execution;
\textbf{Scalability:} Running Docker in the cloud allows for scaling up the number of runners based on demand, ensuring that multiple pipelines can run in parallel without performance degradation;
\textbf{Resource Efficiency:} By pulling only the necessary parts of repositories, we minimize the storage and bandwidth requirements, making the process more efficient.




\subsection{Using RunPod to Run LLaMA Models}
\label{chap:7}

RunPod \cite{runpod2024} is a cost-effective online GPU service offering state-of-the-art GPUs such as the H100.
%
%
We use RunPod's \textit{Serverless} service, which allows the deployment of servers with up to five workers and includes an automatic API request queue management system. Additionally, we employ the \textit{Serverless vLLM} template available on RunPod to quickly deploy servers with any LLM available on Huggingface.
The \textit{Serverless} service operates on a pay-per-use model, charging only when the GPUs are running.


We experimented with several quantized and unquantized models of varying sizes from both the official provider (Meta), Elinas (from Hugging Face), GradientAI, and a research group (NousResearch). 
%
%
%
%
%
After thorough testing (reported in Section \ref{sec:LLaMAmodels}), we finalized the {\small \texttt{gradientai/Llama-3- 70B-Instruct-Gradient-1048k}} model because the model allows for a context length 
of 1048k tokens. This significant context length enables the model to maintain coherence and reason over long sequences of text.
%
%
With this model, the cost required for processing a class of approximately 100 lines with 3 methods is around 4€; this only depends on the GPU usage on RunPod, which is approximately 20 minutes, involving 53 requests to RunPod.

\subsection{Optimization Techniques and Enhancements}
\label{chap:9}

To improve the performance and accuracy of the LLaMA model for test generation, we introduced some post-processing optimization techniques and enhancements that are essential for ensuring the robustness and reliability of the generated JUnit tests. 


One challenge encountered was that LLaMA occasionally provided text responses along with the code, which caused compilation issues. To address this, we implemented a function to extract only the Java code from LLaMA's response, ensuring that non-code text does not interfere with the compilation process. 

There were also cases in which the response from LLaMA was incomplete. When trying to extract code from such responses, the process could get stuck indefinitely. To solve this, we introduced a timeout of 30 seconds, after which the function abandons the current method and moves to the next one, preventing it from getting stuck indefinitely. This timeout mechanism ensures that the system remains responsive and avoids unnecessary delays.

LLaMA's responses sometimes lacked the necessary package statements, leading to compilation errors. To solve this, we implemented a function to check the presence of a package statement in the code. If it is missing, the function adds the package statement using the information saved during the initial extraction of the method content from the main class. 

Compilation errors also arose due to missing import statements. 
%
To fully overcome this problem, we decided to store all the imports in a JSON file until the process is complete. This JSON file acts as a temporary repository for all identified import statements. During the \texttt{mvn test} phase and after merging the tests into the main file, we add imports from this JSON file. This final step ensures that no imports are overlooked and that all dependencies are properly resolved.

LLaMA occasionally produced syntax errors, particularly in long code snippets. To address this, we used \texttt{javac} to compile the code and identify syntax errors through the standard error output. We then implemented a function to fix these syntax errors based on the error messages. This function parses error messages, identifies common syntax issues, and applies corrective actions, such as fixing mismatched braces, correcting variable declarations, and ensuring proper method signatures. 

Despite instructing LLaMA to include the \texttt{@ExtendWith (MockitoExtension.class)} annotation, such annotation was sometimes missing in the response. We created a function to check for the presence of this annotation and manually add it, if necessary. This function scans the generated test class for the annotation and inserts it at the appropriate location if it is missing, ensuring that Mockito-based tests are correctly configured.

\subsection{Feedback Loop Integration}

To improve the accuracy and reliability of the generated tests, we implemented a feedback loop. When a generated test method fails, we provide LLaMA with the specific test method content and the associated error stack trace. LLaMA then refines the test method until it is passed or the maximum number of iterations specified by the user is reached (which is generally 5 iterations). This feedback loop involves capturing detailed error information, feeding it back into the model, and iteratively improving the test generation. This process helps to refine the test methods, making them more robust and ensuring that they accurately validate the intended functionality.

\begin{table*}[t]
\centering
\caption{Comparing LLaMA Models: Test Generation Statistics}
\label{tab:LLaMA-one}
\resizebox{2\columnwidth}{!}{%
\begin{tabular}{c|cccc|cccc|cccc}
\toprule
\multirow{2}[3]{*}{\bf Project} 
& \multicolumn{4}{c}{\bf Generated Tests} 
& \multicolumn{4}{c}{\bf Total Passed} 
& \multicolumn{4}{c}{\bf Overall Coverage} 
\\
\cmidrule(lr){2-5} \cmidrule(lr){6-9} \cmidrule(lr){10-13} 
 & {\texttt{L7B}} & \texttt{L13B} & \texttt{L34B} & \texttt{L70B} & 
{\texttt{L7B}} & \texttt{L13B} & \texttt{L34B} & \texttt{L70B} & {\texttt{L7B}} & \texttt{L13B} & \texttt{L34B} & \texttt{L70B} 
\\
\midrule
Project 1 (28 methods) 
& 0 & 10 & 48 & 90
& 0 & 3 & 15 & 71
& 0\% & 4\% & 29\% & 94\%
\\
Project 2 (24 methods) 
& 0 & 6 & 36 & 50
& 0 & 1 & 8 & 42
& 0\% & 1\% & 21\% & 91\%
\\
Project 3 (11 methods) 
& 0 & 2 & 20 & 45
& 0 & 0 & 5 & 39
& 0\% & 0\% & 26\% & 81\%
\\
MongodbCRUD (12 methods)
& 0 & 3 & 23 & 38
& 0 & 1 & 8 & 29
& 0\% & 2\% & 38\% & 94\%
\\
\bottomrule
\end{tabular}
}
\end{table*}

\begin{table*}[t]
\centering
\caption{Comparing LLaMA Models: Test Refinement Statistics}
\label{tab:LLaMA-two}
\resizebox{2\columnwidth}{!}{%
\begin{tabular}{c|cccc|cccc|cccc}
\toprule
\multirow{2}[3]{*}{\bf Project} 
& \multicolumn{4}{c}{\bf Passed in the 1st iteration} 
& \multicolumn{4}{c}{\bf Passed after the 5th iteration} 
& \multicolumn{4}{c}{\bf Passed after the 10th iteration} 
\\
\cmidrule(lr){2-5} \cmidrule(lr){6-9} \cmidrule(lr){10-13} 
 & {\texttt{L7B}} & \texttt{L13B} & \texttt{L34B} & \texttt{L70B} & 
{\texttt{L7B}} & \texttt{L13B} & \texttt{L34B} & \texttt{L70B} & {\texttt{L7B}} & \texttt{L13B} & \texttt{L34B} & \texttt{L70B} 
\\
\midrule
Project 1 (28 methods) 
& - & 2 & 12 & 36
& - & 3 & 14 & 65
& - & 3 & 15 & 71
\\
Project 2 (24 methods) 
& - & 1 & 6 & 27
& - & 1 & 7 & 40
& - & 1 & 8 & 42
\\
Project 3 (11 methods) 
& - & 0 & 4 & 25
& - & 0 & 5 & 35
& - & 0 & 5 & 39
\\
MongodbCRUD (12 methods)
& - & 0 & 5 & 19
& - & 1 & 7 & 27
& - & 1 & 8 & 29
\\
\bottomrule
\end{tabular}
}
\end{table*}

\subsection{Handling Private Methods}

Testing private methods (or methods that used private methods) has always posed a significant challenge to automated test generation tools: due to encapsulation, these methods cannot be directly mocked. 
However, since code coverage is the main metric for assessing the quality of our tests, it is important to ensure that every line of code is tested, including those of private methods. 

Tools like \textit{Randoop} \cite{Pacheco2007} and \textit{EvoSuite} \cite{Fraser2011} typically bypass this limitation using techniques such as \textit{reflection}\cite{forman2005reflection}, but this can compromise the design principles of object-oriented programming. 
In contrast,
\NAME\ approaches this problem by testing private methods indirectly through their associated public methods, thus ensuring comprehensive coverage while preserving encapsulation.

Specifically, we first examine the given class to identify all private methods. If a private method is present, we extract it and save its content. For non-private methods, when writing tests, we check if the method uses any private method. If it does, we include the content of the private method in the prompt along with the content related to the method under test.
We then instruct LLaMA to generate a test method for the method under test, specifying not to mock the private method, as it is not accessible from the test class. Instead, the original method in the main class, which directly calls the private method, is used for testing.

Repositories are interfaces that abstract database operations, enabling the application to interact with data sources seamlessly. For private methods that utilize repositories, these are mocked to isolate the method's behavior from external dependencies. This ensures that the test focuses solely on the business logic within the private method without invoking actual database operations. The mocks simulate the repository responses as per the test scenario, enabling accurate and efficient testing.
By integrating the private method content into the test generation prompt, we ensure that the context provided to LLaMA is comprehensive and includes all the necessary details. 

\section{Experiments}

Our tool is now in production, so it is not publicly available. However, interested users can request temporary access to the tool for testing purposes, by writing an email to \url{cubetesterAI@pccube.com}; a guide to use it is available at \url{https://github.com/peer-review-234/guide/blob/main/Guide.pdf}.

In all experiments we conduct, the main metrics we consider is
\emph{code coverage}, a key metrics in JUnit testing that measures the percentage of executable code statements that are executed when the test cases are run. It is calculated by dividing the number of executed statements by the total number of statements in the code, providing a simple way to assess how much of the code has been tested.

We first try to understand which is the better LLaMA model, both under code coverage and under other metrics; we discover that the better performances are obtained when using \texttt{gradientai/Llama-3-70B-Instruct-Gradient- 1048k}. Hence, we consider this model to be the reference one for \NAME.
Then, we try to integrate ChaptGPT in place of LLaMA in our tool and see how the results change. It turns out that moving from LLaMA to ChatGPT reduces the ability of \NAME, in terms of generated tests, test passed, and branch code coverage. This fully justifies the use of LLaMA in \NAME\ and, in particular, this is what we use to perform the next experiments
Finally, we compare \NAME\ with some state-of-the-art tools. As the reported comparisons stress, \NAME\ consistently demonstrates  competitive and better performance in terms of code coverage.

\subsection{Comparing LLaMA Models}
\label{sec:LLaMAmodels}

We start by presenting a comparative analysis of different LLaMA models in their ability to generate JUnit tests. This analysis aims to identify the most effective model for automated test generation for \NAME.
The evaluation focuses primarily on code coverage, but also considers the {\em number of tests generated} (i.e., the total number of JUnit tests generated by each model), the {\em number of passed tests} (both initially and after an iterative refinement), and the \emph{total number of methods} in the project for which JUnit tests were generated.

For this comparative analysis, we selected the following four LLaMA models:
\begin{itemize}
    \item \texttt{L7B} (shorthand for \texttt{meta-llama/Llama-2-7b-hf}) is the smallest model, with limited context length and processing capabilities.
    
    \item \texttt{L13B} (shorthand for \texttt{meta-llama/Llama-2-13b- hf}) is a  medium-sized model, with improved context handling over \texttt{L7B}.
    
    \item \texttt{L34B} (shorthand for \texttt{codellama/CodeLlama-34b- hf}) is a  larger model capable of handling more complex scenarios compared to \texttt{L7B} and \texttt{L13B}.
    
    \item \texttt{L70B} (shorthand for  \texttt{gradientai/Llama-3-70B- Instruct-Gradient-1048k}) is the largest model, with extensive context length and processing power, expected to deliver the best performance.
\end{itemize}
The choice of these particular models was driven by the aim of evaluating how the size of the model and the length of the context affect the generation of the test. Choosing such different models offers an evidence on the different benchmarking capabilities and on the trade-offs between computational efficiency, output quality, and context handling. In addition, these models are relevant in practical applications and their compatibility with existing frameworks facilitates effective experimentation.

The evaluation was carried out on three real-world projects. These projects are representative of typical Java applications, featuring numerous private methods, complex interdependencies, and advanced features like Spring Boot's dependency injection. The projects include tightly coupled modules, multi-level method invocations, and extensive interactions across functional components, reflecting the complexity of enterprise-level Java codebases. The analysis was confined to the service modules, which collectively comprised a total of 13 classes; JUnit tests were written and executed solely for these modules. Unfortunately, we cannot share the code of the three projects because these are real software products deployed in our company (PCCube) and the non-disclosure agreement forbids to make them public. To compensate for this limitation, we also used an open-source project named MongodbCRUD,\footnote{\url{https://github.com/Astha-Tiwari02/MongodbCRUD}} that contains 3 classes and 12 methods in total.

The results in terms of generated tests and test refinement are reported in Tables \ref{tab:LLaMA-one} and \ref{tab:LLaMA-two}.
\texttt{L7B} struggled significantly due to its limited context length and processing power, resulting in no generated tests across all projects.
Due to the absence of generated tests, its code coverage was 0\% and no refinement analysis was applicable for it.
\texttt{L13B} showed some improvement over \texttt{L7B}, managing to generate a few test cases, primarily for smaller methods with minimal dependencies.
The code coverage achieved was generally low.
Furthermore, 3 tests passed in the first iteration, 2 additional tests passed after 5 iterations, and no further test passed after 10 iterations.
\texttt{L34B} demonstrated moderate success in generating a substantial number of test cases, showing improvements in both quantity and quality of generated tests.
Furthermore, iterations always provided an increased number of test passed.
\texttt{L70B} outperformed all other models, generating the highest number of tests with significantly better coverage and pass rates.
The code coverage achieved by \texttt{L70B} was notably high: over 95\% for 11 classes, with exceptions with 46\% and 87\% coverage for 2 classes (due to large methods with extensive dependencies). Again, iterated runs lead to a sensible increase of the number of tests passed.

Although \texttt{L70B} provides superior accuracy and coverage, it also incurs a higher cost: for a Java class of approximately 100 lines with three methods, it incurs a cost of around 4€; this cost only depends on the GPU usage on RunPod, which is approximately 20 minutes and involves 53 requests to RunPod. However, also running the smaller LLaMA models involves a cost, of around 2€/3€, and still requires around 20 minutes of work. Thus, while the larger model offers better performance, the cost advantage of using smaller models is relatively modest, suggesting that the investment in the larger model may be justified for achieving better test generation outcomes.

%
%
%
To conclude, \texttt{L70B} demonstrated significant improvement over the other models. Its ability to handle larger context lengths allows it to generate a higher number of comprehensive test cases with superior coverage and pass rates. The model's efficiency in refining tests contributes to its outstanding performance, making it the most effective model in this evaluation.

\begin{table}[t]
\centering
\caption{Comparison between ChatGPT and LLaMA integrated within \NAME\ on 3 Java Projects}
\label{tab:llamavschatgpt}
\resizebox{0.9\columnwidth}{!}{%
\begin{tabular}{c|ccc}
\toprule
 & \textbf{ChatGPT} & \textbf{LLaMA} \\
\midrule
\textbf{Generated Tests} & 168 & 185 \\
\textbf{Total Tests Passed} & 114 & 152 \\
\textbf{Average Coverage} & 76.3\% & 88.6\% \\
\bottomrule
\end{tabular}
}
\end{table}

\begin{table*}[t]
\centering
\caption{Result of experiment: Line coverage comparison across projects, including \NAME\ .}
\label{tab:experiment_results}
\resizebox{2\columnwidth}{!}{%
\begin{tabular}{c|c|c|cccc}
\toprule
\textbf{Project Information} & \textbf{Version} & \textbf{Unseen} & \textbf{ChatUniTest} & \textbf{TestSpark} & \textbf{EvoSuite} & \textbf{\NAME\ } \\
\midrule
\textbf{Cli} & 1.5.0 & no & 70.9\% & 78.4\% & 91.8\% & 91\% \\
\textbf{Csv} & 1.10.0 & no & 73.3\% & - & 28.3\% & 94\% \\
\textbf{Ecommerce} & 695a6d4 & yes & 26.7\% & 36.7\% & - & 73\% \\
\textbf{Binance} & 2.0.0 & yes & 49.2\% & 29.7\% & 20.8\% & 86\% \\
\midrule
\textbf{Overall Coverage} & - & - & 59.6\% & 42.1\% & 38.2\% & 86\% \\
\bottomrule
\end{tabular}
}
\end{table*}

\begin{table*}[t]
\centering
\caption{Line and branch coverage comparison on \textbf{HUMANEVAL} dataset across different tools/models, and \NAME\ .}
\label{tab:humaneval_comparison}
\resizebox{2\columnwidth}{!}{%
\begin{tabular}{c|ccccccc}
\toprule
\textbf{Metric} & \textbf{GPT-3.5-Turbo} & \textbf{StarCoder} & \textbf{Codex-2K} & \textbf{Codex-4K} & \textbf{Evosuite} & \textbf{Manual} & \textbf{\NAME\ } \\
\midrule
Line Coverage & 69.1\% & 67.0\% & 87.4\% & 87.7\% & 96.1\% & 88.5\% & \textbf{ 97.4\% } \\
Branch Coverage & 76.5\% & 69.3\% & 92.1\% & 92.8\% & 94.3\% & 93.0\% & \textbf{ 94.4\% } \\
\bottomrule
\end{tabular}
}
\end{table*}

\subsection{Comparing LLaMA with ChatGPT}

A further experiment consists of integrating ChaptGPT in place of LLaMA in our tool and seeing how the results change.
To ensure a fair and unbiased comparison between LLaMA and ChatGPT, identical prompts were used for both models throughout the experiments. This decision was made to focus the evaluation on the inherent capabilities of the models in generating JUnit tests, rather than on differences introduced by prompt engineering.
We run the two versions of \NAME\ (one with ChatGPT \texttt{3.5 Turbo} and the other with LLaMA \texttt{L70B}) on the first three projects used in Tables \ref{tab:LLaMA-one} and \ref{tab:LLaMA-two}; the outcome is reported in Table \ref{tab:llamavschatgpt}. It turns out that moving from LLaMA to ChatGPT reduces the ability of \NAME\ in terms of generated tests (168 vs 185), in terms of tests passed (114 vs 152), and in terms of code coverage (76.3\% vs 88.6\%). This fully justifies the use of LLaMA \texttt{L70B} in \NAME\ and, in particular, this is what we used for the next experiments.

\subsection{Comparing \NAME\ with State-of-the-Art Tools}

In this section, we compare the performance of \NAME\ in terms of code coverage with several state-of-the-art tools and models for the automated generation of unit tests. We perform three comparisons on different datasets, highlighting the significant improvement achieved by \NAME\ in handling complex codebases and generating comprehensive test suites.
Specifically, we compare \NAME\ with the results reported in \cite{GV23,CHZ24,SNET24} because such papers align closely with the methodologies and goals of our study. These papers provide key insights into the use of LLMs, such as ChatGPT and others, for the automated generation of unit tests, which is a central focus of this paper.
By comparing these papers with our research, we can validate our approach, highlight key differences in methodology, and better understand the general landscape of LLM-based unit test generation.

\subsubsection{Comparison with ChatUniTest \cite{CHZ24}}

The first comparison is made with the tool \texttt{ChatUniTest}, which focuses on generating unit tests using large language models for seen and unseen projects. In their study, \texttt{ChatUniTest} achieved an overall average coverage of 59.6\% on projects\footnote{The projects used for the comparison are available at:
\url{https://github.com/apache/commons-cli} (\textbf{Cli}), 
\url{https://github.com/apache/commons-csv} (\textbf{Csv}),
\url{https://github.com/SelimHorri/ecommerce-microservice-backend-app} (\textbf{Ecommerce}), and 
\url{https://github.com/binance/binance-connector-java} (\textbf{Binance}).
} 
seen and unseen during LLM training (see Table 1 in \cite{CHZ24}). By contrast, \NAME, which preprocesses codebases and iteratively refines tests using error logs, provides a much higher average coverage of 86\% on the same dataset. Table \ref{tab:experiment_results} summarizes the results.


\subsubsection{Comparison with other tools based on LLMs}

Next, we compare our results with those reported in \cite{SST24}. This study tested multiple models, including \texttt{GPT-3.5-Turbo} \cite{gptTurbo}, \texttt{StarCoder} \cite{starcoder}, \texttt{Codex-2K} and \texttt{Codex-4K} \cite{codex}, and \texttt{EvoSuite} \cite{FA16}, on the HumanEval dataset \cite{HumanEval}\footnote{  \url{https://github.com/ASSERT-KTH/human-eval-java }
} (see Table 3 of \cite{SST24}). Our tool achieves the highest line coverage of 97.4\% and branch coverage of 94.4\%, outperforming all the models tested in the study. The detailed results are in Table 
\ref{tab:humaneval_comparison}.


We remark that \cite{SST24} also used the \texttt{SF110} dataset for evaluation. However, we cannot use it because the dataset is outdated and does not include the Spring Boot Framework, which is critical for contemporary Java applications. 
In addition, \cite{SST24} uses other evaluation metrics, such as compilation rates and test smells. In our case, we do not consider compilation rates because our tool automatically refines the generated tests in the event of compilation errors, ensuring that the tests are always compilable. Furthermore, finding test smells was not our focus: our approach prioritizes automated test generation and refinement, aiming for high code coverage and functional accuracy over manual inspection of test smells, which can be subjective and do not directly impact the automation process.

\subsubsection{Comparison with a ChatGPT-based Tool}

Finally, we evaluated \NAME\ against the same programs as those reported in \cite{GV23}, a study that investigated the unit test generation capabilities of ChatGPT on a dataset of 33 Java programs taken from \cite{AR21}\footnote{ \url{https://github.com/aurimrv/initial-investigation-chatgpt-unit-tests}
}. Their tool achieved an overall coverage of 90.2\%. The Baseline Suite used in the study, which includes \texttt{EvoSuite} \cite{FA16}, \texttt{Randoop} \cite{PE07}, \texttt{Palus} \cite{Z11}, and \texttt{JTExpert} \cite{SPG15}, reached a coverage of 99.5\% (see the line AVG of Table 6 in \cite{GV23}). \NAME\ achieves a competitive overall coverage of 97\%, which surpasses the ChatGPT tool and is close to the Baseline Suite, as reported in Table \ref{tab:java_comparison}.

\begin{table}[t]
\centering
\caption{Comparison of Coverage between ChatGPT Tool, Baseline Suite, and \NAME\ on 33 Java Programs}
\label{tab:java_comparison}
\resizebox{\columnwidth}{!}{%
\begin{tabular}{c|ccc}
\toprule
\textbf{Tool/Model} & \textbf{Dataset} & \textbf{Overall Coverage} \\
\midrule
\texttt{ChatGPT Tool} & 33 Java Programs & 90.2\% \\
\texttt{Baseline Suite} & 33 Java Programs & 99.5\% \\
\texttt{\NAME\ } & 33 Java Programs & 97\% \\
\bottomrule
\end{tabular}
}
\end{table}

\subsection{Summary of Comparisons}

In all comparisons, \NAME\ consistently shows a superior, or at least a highly competitive, performance in terms of code coverage. By leveraging preprocessing, handling complex code dependencies, and iteratively refining tests based on error logs, our tool not only generates robust test suites, but also ensures the highest code coverage. These results underline the effectiveness and scalability of our solution in comparison with other state-of-the-art LLM-based unit test generation tools.


\section{Conclusions and Future Work}


This paper explored the application of LLMs, specifically the LLaMA model (and, in particular, its 70B variant), to automate the generation of JUnit tests. Through the integration of LLaMA into a CI/CD pipeline using GitLab and Docker, and the development of a user-friendly web interface. The research produced a tool, named \NAME, that improves the efficiency and accuracy of software testing processes. Additionally, the detailed cost model associated with running LLaMA models in a cloud environment offers valuable insight into the economic feasibility of this approach, making it accessible to a broader range of developers and organizations.
The maturity of \NAME\ has been validated through extensive testing and refinement, making it ready for deployment in production environments. 
Further research can focus on optimizing the LLaMA model for test generation, particularly in reducing the computational resources required, and in improving the speed of the generation; this would make it more accessible for smaller projects or organizations with limited resources.
Moreover, developing collaborative features within the web interface can allow multiple developers to interact with the LLaMA model simultaneously, facilitating team-based approaches to test generation and refinement.
Even more, investigating adaptive learning techniques where the LLaMA model continuously improves its test generation capabilities based on feedback from developers can lead to more and more accurate and relevant test cases.

An orthogonal direction is the enhancement of the feedback loop with more sophisticated error analysis to provide better guidance for test refinement. Incorporating machine learning techniques to analyze error patterns and suggest specific improvements could further enhance the reliability of the generated tests.



We also plan to introduce a test oracle in \NAME\ to validate the correctness of the automatically generated JUnit tests. Although the feedback loop uses error logs to iteratively refine test cases, a formal test oracle—such as specifications or expected outputs—could further enhance the reliability and automation of this process. By integrating such oracles, \NAME\ could systematically verify the alignment of the test output with the expected software behavior, reducing manual intervention, and improving the overall accuracy of the generated test suite.

Future improvements to the web interface include improving the chatbot’s ability to understand more complex user queries, integrating additional testing frameworks, and expanding the interface to support collaborative test generation among multiple developers. These enhancements would contribute to a more robust and versatile tool, capable of meeting the evolving needs of development teams.


Last but not least, extending the research to apply LLaMA models to test generation for other programming languages will broaden the impact and applicability of this approach.

\bibliographystyle{IEEEtran}


\end{document}